# Magnetic resonance in cerium hexaboride caused by quadrupolar ordering

S.V. Demishev[a], A.V. Semeno[a], A.V.Bogach[a], Yu.B. Paderno[b], N.Yu. Shitsevalova[b] and N.E. Sluchanko[a]

[a]*Low Temperatures and Cryogenic Engineering Department, A.M.Prokhorov General Physics Institute of Russian Academy of Sciences, Vavilov street, 38, 119991 Moscow, Russia*

[b]*Department of Refractory Materials, Institute for Problems of Materials Science of Ukranian National Academy of Sciences, Krzhyzhanovsky street, 3, 03680 Kiev, Ukraine*

**Abstract**

Experimental evidence of the magnetic resonance in the antiferro-quadrupole phase of $CeB_6$ is reported. We have shown that below orbital ordering temperature a new magnetic contribution emerge and gives rise to observed resonant phenomenon.

Cerium hexaboride is considered as a typical example of a dense Kondo-system, where $Ce^{3+}$ magnetic ions are arranged in the simple cubic lattice and quadrupole interactions are supposed to play an essential role [1-6]. The current understanding of the genesis of the anomalous properties of $CeB_6$ can be summarized as follows. (1) The interplay between magnetic dipole and electric quadrupole interactions leads to a complicated magnetic phase diagram. In zero magnetic field the quadrupole ordering occurs at $T_Q$=3.2 K and precedes the formation of an antiferromagnetic phase (i.e. dipole ordering) at $T_D$=2.3 K. The application of the external magnetic field induces an enhancement of $T_Q$ and suppression of $T_D$. This sequence of phase transitions in $CeB_6$ has been proved by means of neutron diffraction [1] and resonant X-ray scattering [2] studies as well as by specific heat [3], NMR [4], magnetisation [5-6] and transport measurements [7].

(2) The existence of the quadrupole moment reflects the deviation of the shape of 4*f* shell of Ce ion from spherical, and thus ordering of quadrupoles is equivalent to ordering of *f*-orbitals. The microscopic explanation of the existence of the quadrupole moment is based on the crystal field splitting of the $^2F_{5/2}$ level, which leads to the lowest in energy $\Gamma_8$ term [8]. The neutron scattering evidence [1] of an antiferromagnetic component with a wave vector $\boldsymbol{k}_0$=[½, ½, ½] in the quadrupole phase implies two types of non-equivalent Ce ions at $T < T_Q$ having quadrupole moments $+Q$ and $-Q$ and arranged in an alternating three-dimensional structure. Thus the ordered phase of $CeB_6$ at $T<T_Q$ is referred as an antiferro-quadrupole (AFQ) phase [1-8]. In the region $T > T_Q$ cerium hexaboride is a paramagnetic metal [5-6].

The known theoretical models describing magnetic properties of $CeB_6$ assume that field and temperature dependences of magnetisation $M(B,T)$ originate from a single contribution of localized magnetic moments (LMM) on Ce-ions affected by Kondo screening and quadrupolar effects [9]. In the present work the low temperature magnetic properties of cerium hexaboride have been probed by magnetic resonance and magnetisation measurements. We have observed a novel magnetic resonance in AFQ phase and argue that it’s appearance reflects the emergence in the orbital ordering region of the specific term in $M(T,B)$. The obtained experimental data are inconsistent with the existing dense Kondo-system and quadrupolar models for $CeB_6$ and therefore these theoretical models require clarification.

When studying magnetic resonance we have measured the transmission of the copper cylindrical cavities operating at $TE_{01n}$ modes and tuned to frequencies 60, 78 and 100 GHz. As a source of microwave radiation backward wave oscillators have been used. One of the endplates of the cylinder has been made of the high quality $CeB_6$ single crystal, and the holes connecting cavity to waveguides were located at the other endplate. The cavity quality factor was about ~$6 \cdot 10^3$. The cavity has been placed in the cryostat with a 7 T superconducting magnet and the experimental setup has allowed controlling the cavity temperature with an accuracy better than 0.01 K down to 1.7 K. The magnetic field was aligned along the [110] crystallographic direction and was parallel to the cavity axis. As a reference a small DPPH sample has been placed in the cavity. The quality of the $CeB_6$ single crystal has been controlled by means of X-ray and chemical analysis as well as by transport and magnetic measurements.

The obtained data are summarised in fig. 1. For $T > T_Q(B)$ the field dependence of the cavity transmission displays a bend, which shifts to lower magnetic field when temperature is lowered (fig. 1). This feature reflects the change of the $CeB_6$ impedance at the boundary between paramagnetic and antiferro-quadrupole phases [10].

In contrast to the paramagnetic phase, for $T<T_Q$ a new strong resonance develops and the intensity of this line increases with lowering temperature (fig. 1). Frequency measurements have shown that the position of the observed line shifts linearly corresponding to a Lande $g$-factor 1.62, which strongly deviates from the values $g$=1.98-2.5 reported for the EPR in the paramagnetic phase [8]. At low temperatures the bend of the transmission curves corresponding to the boundary between antiferromagnetic and antiferro-quadrupole phase has been also detected (fig. 1).

Note that the positions of the bends in fig. 1 completely agree with magnetic phase diagram subtracted from the magnetization and magnetoresistance experiments, including data obtained in the present work as well as reported previously [5-7]. Detailed discussion of the magnetic phase diagram is given elsewhere [11].

In described experiment both bulk and surface properties of the sample may contribute to a cavity microwave response. Therefore we have undertaken special measures to exclude an explanation of the observed resonance by a surface effect. For this purpose various kinds of the sample surface preparations have been checked, including several roughnesses of the mechanical polishing and different regimes of chemical etching. Experiments with various surface treatments have provided results identical to those presented above; the excitation of the resonant mode below $T_Q$ has been also checked for several samples. Note that observation of the phase boundaries subtracted from the volume properties like magnetisation and resistivity in the same cavity experiment also favours the understanding of the observed resonance as a bulk effect.

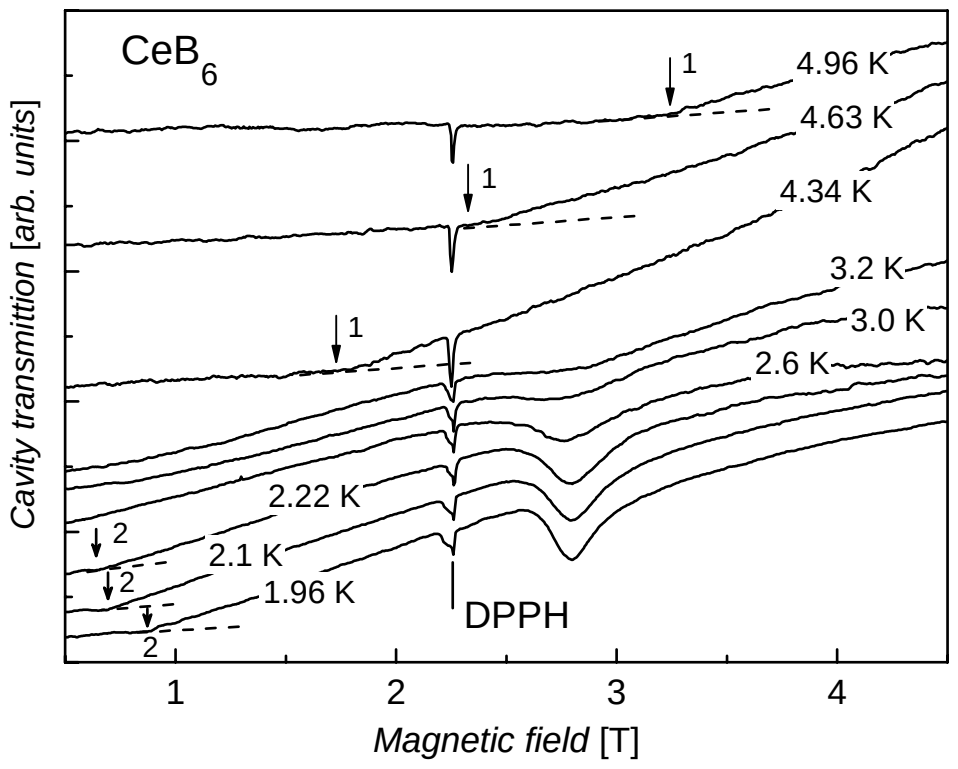

Fig. 1. Field dependences of cavity transmission at various temperatures. Arrows mark magnetic phase transitions: (1) AFQ phase - paramagnetic phase; (2) antiferromagnetic phase - AFQ phase.

In order to clarify the nature of magnetic moment, which oscillations give rise to the resonance, we have measured temperature and field dependences of magnetisation of the same crystals in the domain 1.8-50 K/5 T using vibrating sample magnetometer. As long as the resonant field $B_{res}$(60 GHz)=2.8 T is temperature independent (fig. 1) it is worth to compare temperature dependences of magnetisation $M(B_{res},T)$ and integrated intensity $I(T)$ for the resonant mode. The data in fig. 2 show that $I(T)$ does not follow $M(B_{res},T)$ as it should in case [9], when magnetic moment of $CeB_6$ originates from a single "source", such as $Ce^{3+}$ magnetic ions.

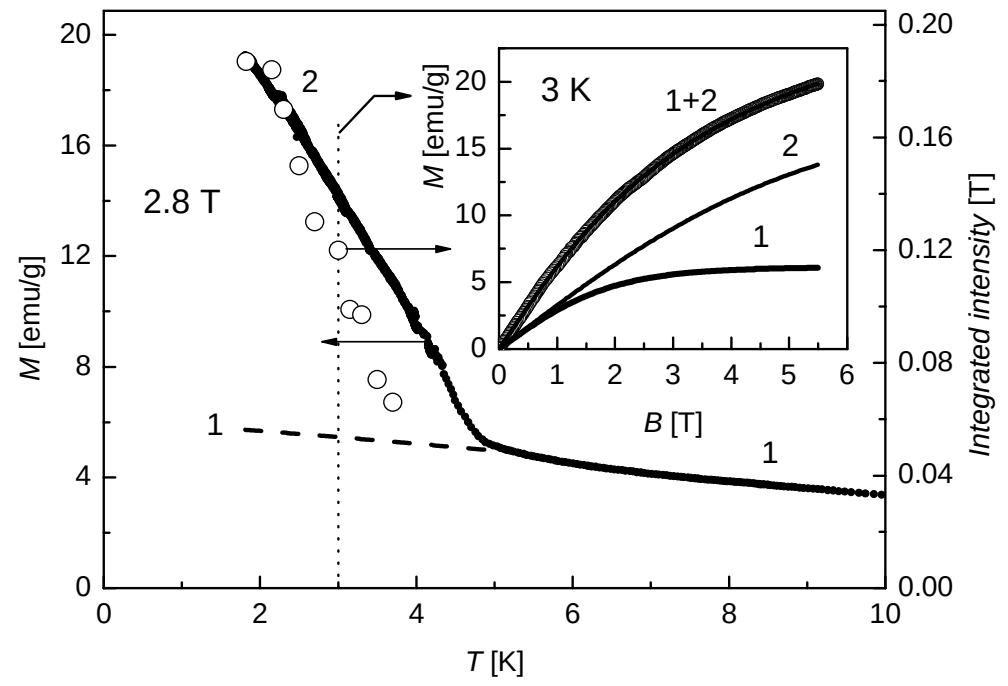

Fig. 2. Temperature dependence of the integrated intensity and magnetisation in resonant field $B_{res}$. Inset shows field dependence of magnetisation at 3 K (dotted line in the main figure). At inset points denote experimental data, solid line represents best fit with Eq. (1). Curves 1 and 2 are calculated magnetic contributions $M_1$ and $M_2$. Dashed line in the main figure corresponds to estimated "base line" (see text for details).

Therefore we assumed that $M(B,T)$ can be represented as a superposition of two terms

$$M(B,T) = M_1(B,T) + M_2(B,T) = A_1 \cdot \tanh\left(\frac{\mu_1 B}{k_B (T-\Theta)}\right) + A_2 \cdot \tanh\left(\frac{\mu_2 B}{k_B T}\right) \qquad (1)$$

In Eq. (1) the first term $M_1(B,T)$ exists in the whole temperature range, while the second one develops at $T < T_Q$ and $M_2(B,T)=0$ for $T > T_Q$. The expression for $M_2(B,T)$ is supposed to model magnetisation of LMM responsible for magnetic resonance; as long as $g=const$ in resonant experiment (fig. 1) the magnetic moment magnitude $\mu_2$ should not depend on magnetic field and temperature and therefore may be estimated as $\mu_2=g\mu_B M=0.81\mu_B$ assuming effective quantum number $M$=1/2. The analytic form for the "base line" $M_1(B,T)$ was chosen to represent Curie-Weiss susceptibility with $\Theta$=-4.5 K found in weak magnetic field and saturation of magnetisation in strong magnetic field.

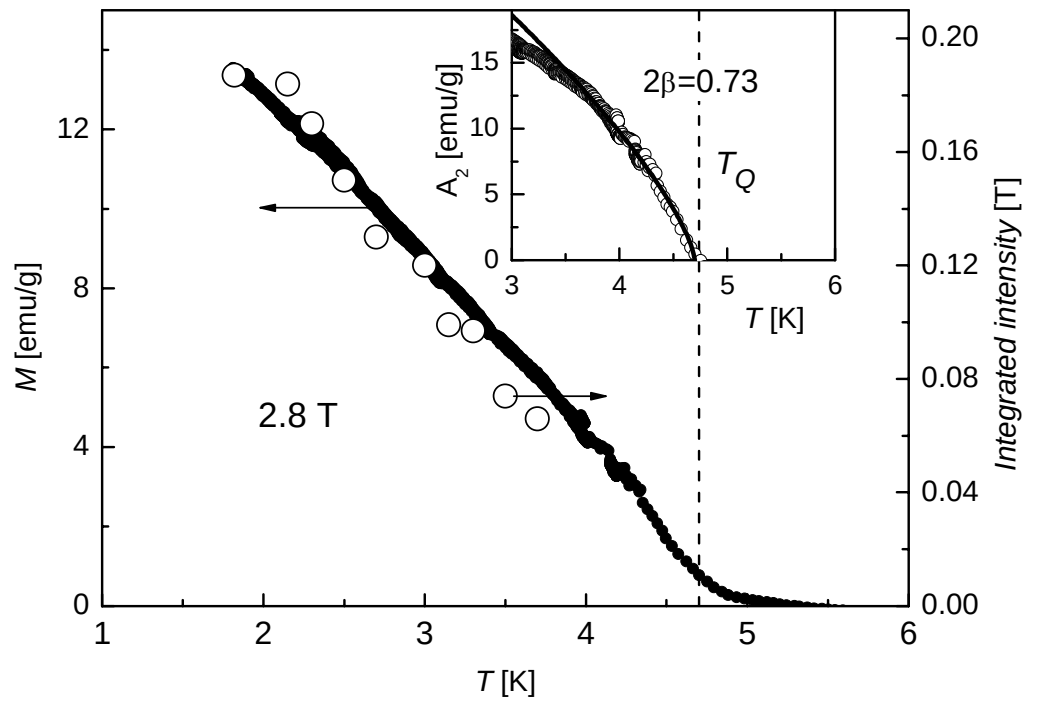

Fig. 3. Comparison of the subtracted contribution $M_2(B_{res},T)$ and integrated intensity for $T<T_Q$. Inset shows obtained temperature dependence of $A_2$ parameter (points) and calculated critical behaviour (solid line) with the critical exponent $2\beta$.

The remaining parameters in Eq. (1) can be found from the field dependence of magnetisation. The curve $M(B, T=3\text{ K})$, which is not affected by abrupt changes caused by phase transitions, have been used to find $A_1$, $A_2$ and $\mu_1$. The inset in fig. 2 shows result of field dependence modelling together with $M_1(B,T)$ and $M_2(B,T)$ terms providing best fit (curves 1 and 2 respectively). It is visible that Eq.(1) describes well experimental data. We have used calculated $M_1(B=B_{res},T)$ to extrapolate $M(B_{res},T)$ in the diapason $T<T_Q$ assuming $A_1=const$ (dashed curve 1 in fig. 2) and, finally, subtract contribution related with the magnetic resonance.

Fig. 3 elucidates that subtracted term $M_2(B_{res},T)$ reproduces well temperature dependence of the integrated intensity. Therefore it is possible to conclude that suggested procedure of separation of different contributions to magnetisation allows describing simultaneously the characteristics of the oscillated magnetic moment, field and temperature dependences of magnetisation and temperature dependence of the integrated intensity.

Interesting that fitting procedure suggests relatively high magnetic moments $\mu_1 \approx 5.7\mu_B$ and critical behaviour of $A_2(T) \sim (T_Q-T)^{2\beta}$ (inset in fig. 3). The observed critical exponent $2\beta=0.73\pm0.01$ practically coincide with the index $2\beta=0.74$ describing temperature dependence of intensity for the reflex $\boldsymbol{k}_0$=[½, ½, ½], which is specific to AFQ phase [2].

Therefore first term in Eq.(1) may originate from the magnetism of the strongly correlated quasi particles (heavy fermions) with renormalized $\mu$ whereas the second term reflects LMM caused by orbital ordering at $T < T_Q$. Although this hypothesis agrees with some recent neutron scattering experiments [12] it requires further theoretical and experimental investigation.

Summarising, we have shown that below quadrupole ordering temperature $T<T_Q$ a new magnetic contribution develops and gives rise to observed magnetic resonance. This behaviour is hardly possible to expect in dense Kondo system, where LMM should vanish al low temperatures rather than emerge. From the other hand, in the quadrupole ordering concept, where magnetism of Ce magnetic ions is solely accounted, is difficult to explain splitting of magnetisation into two components having different physical nature. Therefore an adequate theory explaining magnetic properties of $CeB_6$ including magnetic resonance and orbital ordering appears on the agenda.

**Acknowledgements**

This work was supported by the INTAS project 03-51-3036 and by the program of the Russian Academy of Sciences “Strongly Correlated Electrons”. Part of the work was supported by RFBR grants 04-02-16574 and 04-02-16721.

**References**

[1] J. M.Effantin et al., J.Magn. and Magn. Materials. 47&48 (1985) 145.
[2] H.Nakao et al., J. Phys. Soc. Jpn. 70 (2001) 1857.
[3] K.N.Lee and B.Bell, Phys. Rev. B. 6 (1972) 1032.
[4] M.Takigawa et al., J. Phys. Soc. Jpn. 52 (1983) 728.
[5] M.Kawakami et al., Solid State Commun. 36 (1980) 435.
[6] D.Hall, Z.Fisk and R.G.Goodrich, Phys.Rev.B.62 (2000) 84.
[7] A.Takase et al., Solid State Commun. 36 (1980) 461
[8] C.Terzioglu et al., Phys. Rev. B. 63 (2001) 235110.
[9] M.Sera, Y.Kobayashi, J. Phys. Soc. Jpn. 68 (1999) 1664
[10] N.E.Sluchanko et al., JETP Lett. 63 (1996) 453.
[11] S.V.Demishev et al., phys. stat. sol. (b), 242 (2005) R27
[12] V.P.Plakhty et al., Phys. Rev. B, 71 (2005) 100407.